\documentclass{article}[12pt]
\usepackage{amsmath,amsthm,amssymb,amsfonts,amscd,eucal}

\numberwithin{equation}{section}

\def\bn{{\mathbb N}}

\def\itm#1{\item{$(#1)$}}

\DeclareMathOperator{\sgn}{sgn}

\newtheorem{theorem}{Theorem}[section]

\newtheorem{proposition}[theorem]{Proposition}
\newtheorem{corollary}[theorem]{Corollary}
\newtheorem{definition}{Definition}[section]
\newtheorem{example}{Example}[section]
\newtheorem{lemma}[theorem]{Lemma}
\theoremstyle{definition}
\newtheorem{remark}{Remark}[section]

\makeindex

\begin{document}

\title{A Robertson-type Uncertainty Principle and Quantum Fisher Information}

\author{
Paolo Gibilisco\footnote{Dipartimento SEFEMEQ and Centro V.Volterra, Facolt\`a di
Economia, Universit\`a di Roma ``Tor Vergata", Via Columbia 2, 00133
Rome, Italy.  Email:  gibilisco@volterra.uniroma2.it --
URL: http://www.economia.uniroma2.it/sefemeq/professori/gibilisco},
Daniele Imparato\footnote{Dipartimento di Matematica, Politecnico di Torino,
Corso Duca degli Abruzzi 24, 10129 Turin, Italy.
Email: daniele.imparato@polito.it}
\
and
Tommaso Isola\footnote{Dipartimento di Matematica,
Universit\`a di Roma ``Tor Vergata",
Via della Ricerca Scientifica, 00133 Rome, Italy.
Email: isola@mat.uniroma2.it --
URL: http://www.mat.uniroma2.it/$\sim$isola}
}

\maketitle

\begin{abstract}

Let $A_1,...,A_N$ be complex selfadjoint matrices and let $\rho$ be a density matrix. The Robertson uncertainty principle
$$
{\rm det}\left\{ {\rm Cov}_{\rho}(A_h,A_j) \right\} \geq {\rm det} \left\{ - \frac{i}{2} {\rm Tr}(\rho [A_h,A_j])\right\}
$$
gives a bound for the quantum generalized covariance in terms of the commutators $ [A_h,A_j]$. The right side matrix is antisymmetric and therefore the
bound is trivial (equal to zero) in the odd case $N=2m+1$.

Let $f$ be an arbitrary normalized symmetric operator monotone function and let $\langle \cdot, \cdot \rangle_{\rho,f}$ be the associated quantum Fisher information.
In this paper we prove the inequality
$$
{\rm det}\left\{ {\rm Cov}_{\rho}(A_h,A_j) \right\} \geq {\rm det} \left\{ \frac{f(0)}{2} \langle i[\rho, A_h],i[\rho,A_j] \rangle_{\rho,f}\right\}
$$
that gives a non-trivial bound for any $N \in {\mathbb N}$ using the commutators $[\rho,A_h]$.
\smallskip

\noindent 2000 {\sl Mathematics Subject Classification.} Primary 62B10, 94A17; Secondary 46L30, 46L60.

\noindent {\sl Key words and phrases.} Generalized variance, uncertainty principle, operator monotone functions, matrix means, quantum Fisher information.

\end{abstract}

\section{Introduction}

Let $(V, g(\cdot,\cdot))$ be a  real inner-product  vector space and
suppose that $v_1, ... , v_N \in V$. The real $N \times N$ matrix
$G:=\{g(v_h,v_j) \}$ is positive semidefinite and one can define
${\rm Vol}^g(v_1, ... ,
v_N):=\sqrt{{\rm det}\{g(v_h,v_j) \}}$. If the inner product depends
on a further parameter in such a way that
$g(\cdot,\cdot)=g_{\rho}(\cdot,\cdot)$, we write ${\rm Vol}^g(v_1,
... , v_N)={\rm Vol}_{\rho}^g(v_1, ... , v_N)$.

As an example, consider a probability space $(\Omega, {\cal
G},\rho)$ and let $V={\cal L}^2_{\mathbb R}(\Omega, {\cal G},\rho)$
be the space of square integrable real random variables endowed with
the scalar product given by the covariance ${\rm
Cov}_{\rho}(A,B):={\rm E}_{\rho}(AB)-{\rm E}_{\rho}(A){\rm
E}_{\rho}(B)$. For $A_1, ..., A_n \in {\cal L}^2_{\mathbb R}(\Omega,
{\cal G},\rho)$, $G$ is the well known covariance matrix and one has
\begin{equation}
{\rm Vol}_{\rho}^{\rm Cov}(A_1, ... , A_N) \geq 0. \label{gvar}
\end{equation}
The expression ${\rm det} \{{\rm Cov}_{\rho}(A_h,A_j\}$ is known as
the {\sl generalized variance} of the random vector $(A_1,...,A_N)$ and,
in general, one cannot expect a stronger inequality. For instance,
when $N=1$,  $(\ref{gvar})$ just reduces to ${\rm Var}_{\rho}(A)
\geq 0$.

In non-commutative probability the situation is quite different due to the possible non-triviality of the commutators $[A_i,A_j]
$. Let  $M_{n,sa}:=M_{n,sa}(\mathbb{C})$ be the space of all $n
\times n$ self-adjoint matrices and let ${\cal D}_n^1$ be the set of
strictly positive density matrices (faithful states). For $A,B \in
M_{n,sa}$ and $\rho \in {\cal D}_n^1$ define the (symmetrized)
covariance as ${\rm Cov}_{\rho}(A,B):=1/2[{\rm Tr}(\rho A B)+{\rm
Tr}(\rho B A)]-{\rm Tr}(\rho A)\cdot{\rm Tr}(\rho B)$. If $A_1, ...
, A_N$ are self-adjoint matrices  one has
\begin{equation}
{\rm Vol}_{\rho}^{\rm Cov}(A_1, ... , A_N) \geq \begin{cases}
    0, & N=2m+1,\\
    {\rm det} \{-\frac{i}{2}{\rm Tr}(\rho [A_h,A_j])\}^{\frac{1}{2}}, & N=2m.
\end{cases}\label{nup}
\end{equation}
Let us call (\ref{nup}) the ``standard" uncertainty principle to
distinguish it from other inequalities like the ``entropic"
uncertainty principle and similar inequalities. Inequality
(\ref{nup}) is due to Heisenberg, Kennard, Robertson and
Schr\"odinger for $N=2$ (see
\cite{Heisenberg:1927}\cite{Kennard:1927} \cite{Robertson:1929}
\cite{Schroedinger:1930}). The general case is due to Robertson (see
\cite{Robertson:1934}). Examples of recent references where
inequality (\ref{nup}) plays a role are given by
\cite{Trifonov:1994} \cite{Trifonov:2000} \cite{Trifonov:2002}
\cite{DodonovDodonovMizrahi:2005} \cite{Daoud:2006}
\cite{JarvisMorgan:2006}.

We are not aware of any general inequality of type (\ref{nup})
giving a bound also in the odd case $N=2m+1$. If one considers the
case $N=1$, it is natural to seek such an inequality in terms of the
commutators $[\rho,A_i]$.

The purpose of the present paper is to proof  an inequality similar to (\ref{nup}) but not trivial for
any $N \in {\mathbb N}$. Let ${\cal F}_{op}$ be the family of
symmetric normalized operator monotone functions. To each element $f
\in {\cal F}_{op}$ one may associate a $\rho$-depending scalar
product $\langle \cdot , \cdot \rangle_{\rho,f}$ on the self-adjoint
(traceless) matrices, which is a quantum version of the Fisher
information (see \cite{Petz:1996}). Let us denote the associated
volume by ${\rm Vol}_{\rho}^f$. We shall prove that for any $N \in
{\mathbb N}^+$ (this is one of the main differences from (1.2)) and for
arbitrary self-adjoint matrices $A_1, ... , A_N$ one has
\begin{equation}
{\rm Vol}_{\rho}^{\rm Cov}(A_1, ... , A_N) \geq
\left(\frac{f(0)}{2}\right)^{\frac{N}{2}}{\rm
Vol}_{\rho}^{f}(i[\rho,A_1], ... , i[\rho,A_N]). \label{conj}
\end{equation}

The cases $N=1,2,3$ of inequality (\ref{conj})
have been proved by the joint efforts of a number of authors in
several papers: S. Luo, Q. Zhang, Z. Zhang  (\cite{Luo:2000} \cite{Luo:2003b}
\cite{LuoZZhang:2004} \cite{LuoQZhang:2004} \cite{LuoQZhang:2005}); H. Kosaki
(\cite{Kosaki:2005}); K. Yanagi, S. Furuichi, K. Kuriyama (\cite{YanagiFuruichiKuriyama:2005}); F. Hansen
(\cite{Hansen:2006b}); P. Gibilisco, D. Imparato, T. Isola (\cite{GibiliscoIsola:2007} \cite{GibiliscoImparatoIsola:2007}\cite{GibiliscoImparatoIsola:2007b}).

The scheme of the paper is as follows. In Section \ref{prel} we
describe the preliminary notions of operator monotone functions,
matrix means and quantum Fisher information. In Section \ref{tilde}
we discuss a correspondence between regular and non-regular operator
monotone functions that is needed in the sequel. In Section
\ref{con} we state our main result, namely the inequality
(\ref{conj});  we also state other two results concerning how the
right side depends on $f \in {\cal F}_{op}$ and the conditions to have equality in (\ref{conj}).
 In Section \ref{m} we prove the main results. In Section \ref{confronto} we compare the standard uncertainty principle with inequality (1.3). In Sections \ref{comb}, \ref{H} and \ref{K} we prove some auxiliary results.

\section{Operator monotone functions, matrix means and quantum Fisher information} \label{prel}

Let $M_n:=M_n(\mathbb{C})$ (resp. $M_{n,sa}:=M_{n,sa}(\mathbb{C})$)
be the set of all $n \times n$ complex matrices (resp.  all $n
\times n$ self-adjoint matrices).  We shall denote general matrices
by $X,Y,...$ while letters $A,B,...$ will be used for self-adjoint
matrices, endowed with the Hilbert-Schmidt scalar product $\langle
A,B \rangle={\rm Tr}(A^*B)$.  The adjoint of a matrix $X$ is denoted
by $X^{\dag}$ while the adjoint of a superoperator $T:(M_n,\langle
\cdot,\cdot \rangle) \to (M_n ,\langle \cdot,\cdot \rangle)$ is
denoted by $T^*$. Let ${\cal D}_n$ be the set of strictly positive
elements of $M_n$ and ${\cal D}_n^1 \subset {\cal D}_n$ be the set
of strictly positive density matrices, namely $ {\cal D}_n^1=\{\rho
\in M_n \vert {\rm Tr} \rho=1, \, \rho>0 \} $. If it is not otherwise
specified, from now on we shall treat the case of faithful states, namely
$\rho>0$.

A function $f:(0,+\infty)\to
\mathbb{R}$ is said {\it operator monotone (increasing)} if, for any
$n\in \bn$, and $A$, $B\in M_n$ such that $0\leq A\leq B$, the
inequalities $0\leq f(A)\leq f(B)$ hold.  An operator monotone
function is said {\it symmetric} if $f(x)=xf(x^{-1})$ and {\it
normalized} if $f(1)=1$.

\begin{definition}
${\cal F}_{op}$ is the class of functions $f: (0,+\infty) \to (0,+\infty)$ such that

\itm{i} $f(1)=1$,

\itm{ii} $tf(t^{-1})=f(t)$,

\itm{iii} $f$ is operator monotone.
\end{definition}

\begin{example}
    Examples of elements of ${\cal F}_{op}$ are given by the following
list
\[\begin{array}{rcllrcl}
f_{RLD}(x)&:=&\frac{2x}{x+1},&&
f_{WY}(x)&:=&\left(\frac{1+\sqrt{x}}{2}\right)^2,\\[12pt]
f_{SLD}(x)&:=&\frac{1+x}{2},&& f_{WYD(\beta)}(x)&:=& \beta (1-
\beta) \frac{(x-1)^2}{(x^{\beta}-1) (x^{1-\beta}-1)},\qquad \beta
\in \Bigl(0,\frac{1}{2}\Bigr).
$$\end{array}\]

\end{example}

 We now report Kubo-Ando theory of matrix means (see
\cite{KuboAndo:1979/80}) as exposed in \cite{PetzTemesi:2005}.

\begin{definition}
 A {\sl mean} for pairs of positive matrices is a function
$m:{\cal D}_n \times {\cal D}_n \to {\cal D}_n$ such that

(i) $m(A,A)=A$,

(ii) $m(A,B)=m(B,A)$,

(iii) $A <B  \quad \Longrightarrow \quad A<m(A,B)<B$,

(vi) $A<A', \quad B<B' \quad \Longrightarrow \quad m(A,B)<m(A',B')$,

(v) $m$ is continuous,

(vi) $Cm(A,B)C^* \leq m(CAC^*,CBC^*)$, for every $ C \in M_n$.
\end{definition}

Property $(vi)$ is known as the transformer inequality. We denote by
$\displaystyle {\cal M}_{op}$ the set of matrix means. The
fundamental result, due to Kubo and Ando, is the following.

\begin{theorem}
There exists a bijection between ${\cal M}_{op}$ and ${\cal F}_{op}$ given by
the formula
$$
m_f(A,B):= A^{\frac{1}{2}}f(A^{-\frac{1}{2}} B
A^{-\frac{1}{2}})A^{\frac{1}{2}}.
$$
\end{theorem}

\begin{example}
The  arithmetic, geometric and harmonic (matrix) means are given respectively by
\[\begin{array}{rcl}
A \nabla B&:=&\frac{1}{2}(A+B),\\[12pt]
A\# B&:=&A^{\frac{1}{2}}(A^{-\frac{1}{2}} B
A^{-\frac{1}{2}})^{\frac{1}{2}}A^{\frac{1}{2}}, \\[12pt]
A{\rm !}B&:=&2(A^{-1}+B^{-1})^{-1}.\end{array}\]
They correspond respectively to the operator monotone functions
$\frac{x+1}{2},\sqrt{x},\frac{2x}{x+1}$.
\end{example}

Kubo and Ando  \cite{KuboAndo:1979/80} proved that, among matrix means, arithmetic is the largest while harmonic is the
smallest.

\begin{corollary} \label{basic}
For any $f \in {\cal F}_{op}$ and for any $x,y>0$ one has
$$
\frac{2x}{1+x}\leq f(x) \leq \frac{1+x}{2},
$$
$$
\frac{2xy}{x+y}\leq m_f(x,y) \leq \frac{x+y}{2}.
$$
\end{corollary}

In what follows, if ${\cal N}$ is a differential manifold we denote
by $T_{\rho} \cal N$ the tangent space to $\cal N$ at the point
$\rho \in {\cal N}$.  Recall that there exists a natural
identification
 of $T_{\rho}{\cal D}^1_n$ with the space of self-adjoint traceless
 matrices; namely, for any $\rho \in {\cal D}^1_n $
$$
T_{\rho}{\cal D}^1_n =\{A \in M_n|A=A^* \, , \, \hbox{Tr}(A)=0 \}.
$$

 A Markov morphism is a completely positive and trace preserving operator $T:
M_n \to M_m$. A {\sl monotone metric} is a family of Riemannian metrics $g=\{g^n\}$
 on $\{{\cal D}^1_n\}$, $n \in \mathbb{N}$, such that
 $$
 g^m_{T(\rho)}(TX,TX) \leq g^n_{\rho}(X,X)
 $$
 holds for every Markov morphism $T:M_n \to M_m$, for every $\rho \in
 {\cal D}^1_n$ and for every $X \in T_\rho {\cal D}^1_n$.
Usually monotone metrics are normalized in such a way that
$[A,\rho]=0$ implies $g_{\rho} (A,A)={\rm Tr}({\rho}^{-1}A^2)$.
A monotone metric is also said a {\sl quantum Fisher information} (QFI) because of Chentsov uniqueness theorem for commutative monotone metrics (see \cite{Chentsov:1982}).

Define $L_{\rho}(A):= \rho A$, and $R_{\rho}(A):= A\rho$, and observe
 that they are commuting self-adjoint (positive) superoperators on $M_{n,sa}$. For any $f\in {\cal F}_{op}$ one can define the positive superoperator
$m_f(L_{\rho},R_{\rho})$.
Now we can state the fundamental theorem about monotone metrics.

\begin{theorem} (see \cite{Petz:1996})

    There exists a bijective correspondence between monotone metrics (quantum Fisher informations)
    on ${\cal D}^1_n$ and normalized symmetric operator monotone
    functions $f\in {\cal F}_{op}$.  This correspondence is given by
    the formula
    $$
   \langle A,B \rangle_{\rho,f}:={\rm Tr}(A\cdot
    m_f(L_{\rho},R_{\rho})^{-1}(B)).
    $$
\end{theorem}

The metrics associated with the
functions $f_{\beta}$
are very important in
information geometry and are related to Wigner-Yanase-Dyson
information (see for example \cite{GibiliscoIsola:2001}
\cite{GibiliscoIsola:2003} \cite{GibiliscoIsola:2004}
\cite{GibiliscoIsola:2005} \cite{GibiliscoIsola:2007}  \cite{GibiliscoImparatoIsola:2007} and references therein).

\section{The function $\tilde f$ and its properties} \label{tilde}

For $f \in {\cal F}_{op}$ define $f(0):=\lim_{x\to 0} f(x)$.
The condition $f(0)\not=0$ is relevant because it is a necessary and
sufficient condition for the existence of the so-called radial
extension of a monotone metric to pure states (see
\cite{PetzSudar:1996}).
Following \cite{Hansen:2006b} we say that a function $f \in {\cal
F}_{op}$ is {\sl regular} iff $f(0) \not= 0$.  The corresponding
operator mean, associated QFI, etc.  are said regular
too.

\begin{definition}We introduce the sets
$$
{\cal F}_{op}^{\, r}:=\{f\in {\cal F}_{op}| \quad f(0) \not= 0 \},  \quad
{\cal F}_{op}^{\, n}:=\{f\in {\cal F}_{op}| \quad f(0) = 0 \}.
$$
\end{definition}

Trivially one has ${\cal F}_{op}={\cal F}_{op}^{\, r}\dot{\cup}{\cal F}_{op}^{\, n}$.

\begin{proposition} {\rm \cite{GibiliscoImparatoIsola:2007}}
For $f \in {\cal F}_{op}^{\, r}$ and $x>0$ set
$$
\tilde{f}(x):=\frac{1}{2}\left[ (x+1)-(x-1)^2 \frac{f(0)}{f(x)}
\right].
$$
Then ${\tilde f} \in {\cal
F}_{op}^{\, n}$.
\end{proposition}

By the very definition one has the following result (see Proposition
5.7 in \cite{GibiliscoImparatoIsola:2007}).

\begin{proposition} \label{min}Let $f \in {\cal F}^r_{op}$. The following three conditions are equivalent:

1) $\qquad \tilde{f} \leq \tilde{g}$;

2) $\qquad m_{\tilde{f}} \leq m_{\tilde{g}}$;

3) $\qquad \frac{f(0)}{f(t)} \geq \frac{g(0)}{g(t)} \qquad \forall t >0$.

\end{proposition}

Let us give some more definitions.
\begin{definition}
Suppose that $\rho \in {\cal D}_n^1$ is fixed.  Define $X_0:=X-{\rm
Tr}(\rho X) I$.
\end{definition}
\begin{definition} For $A_1,A_2 \in M_{n,sa}$ and $\rho \in {\cal
D}_n^1$ define covariance and variance as
$$ {\rm
Cov}_{\rho}(A_1,A_2):=\frac{1}{2}[{\rm Tr}(\rho A_1 A_2)+{\rm Tr}(\rho
A_2A_1)]-{\rm Tr}(\rho A_1)\cdot{\rm Tr}(\rho A_2)=
$$
$$
= \frac{1}{2}[{\rm
Tr}(\rho (A_1)_0 (A_2)_0)+{\rm Tr}(\rho  (A_2)_0 (A_1)_0)]={\rm Re}\{{\rm Tr}(\rho
(A_1)_0 (A_2)_0)\},\label{cov}
$$
$$
{\rm Var}_{\rho}(A):={\rm Cov}_{\rho}(A,A)={\rm Tr}(\rho A^2)-{\rm
Tr}(\rho A)^2 = {\rm Tr}(\rho A^2_0).
$$
\end{definition}
Suppose, now, that $A_1,A_2 \in M_{n,sa}$, $\rho \in {\cal D}^1_n$ and
$f \in {\cal F}^r_{op}$.
The fundamental theorem for our present purpose is given by
Proposition 6.3 in \cite{GibiliscoImparatoIsola:2007}, which is stated as
follows.

\begin{theorem} \label{!}
$$
\frac{f(0)}{2} \langle i[\rho,A_1], i[\rho,A_2] \rangle_{\rho,f}={\rm Cov}_{\rho}(A_1,A_2)-{\rm Tr}(m_{\tilde f}(L_{\rho},R_{\rho})((A_1)_0)(A_2)_0).
$$
\end{theorem}

As a consequence of the spectral theorem and of Theorem \ref{!} one
has the following relations.

\begin{proposition}{\rm \cite{GibiliscoImparatoIsola:2007}} \label{?}
Let $\left\{\varphi_i\right\}$ be a complete orthonormal base
composed of eigenvectors of $\rho$, and $\{ {\lambda}_i \}$ the
corresponding eigenvalues. To self-adjoint matrices $A_1$, $A_2$ we
associate matrices ${\cal A}^j= {\cal A}^j(\rho)$ $j=1,2$ whose entries are given
respectively by ${\cal A}^j_{kl} \equiv \langle {(A_j)_0} {\varphi}_k
|{\varphi}_l \rangle $.

We have the following identities.

\begin{align*}
    {\rm Cov}_{\rho} (A_1,A_2)&= {\rm
    Re} \{{\rm Tr} (\rho (A_1)_0 (A_2)_0) \}= \frac{1}{2}
    \sum_{k,l}({\lambda}_k+{\lambda}_l) {\rm Re} \{{\cal A}^1_{kl}{\cal A}^2_{lk} \} \\
    \frac{f(0)}{2}\langle i[\rho,A_1], i[\rho,A_2]\rangle_{\rho,f}
    &= \frac{1}{2} \sum_{k,l}({\lambda}_k+{\lambda}_l) {\rm Re} \{
    {\cal A}^1_{kl}{\cal A}^2_{lk} \} - \sum_{k,l} m_{\tilde f}(\lambda_i,\lambda_j)
    {\rm Re} \{ {\cal A}^1_{kl}{\cal A}^2_{lk}  \} .
\end{align*}\label{fin}
\end{proposition}

We also need the following result (Corollary 11.5 in \cite{GibiliscoImparatoIsola:2007}).

\begin{proposition} \label{puri}
On pure states
$$
{\rm Tr}(m_{\tilde f}(L_{\rho},R_{\rho})((A_1)_0)(A_2)_0)=0.
$$
\end{proposition}

\section{Volume theorems for quantum Fisher informations} \label{con}

If we have a matrix $A=\{A_{kl}\}$ we write for the determinant ${\rm det}(A)={\rm det}\{A_{kl}\}$.

Let $(V, g(\cdot,\cdot))$ be a  real inner-product  vector space. By
$\langle u,v \rangle$ we denote the standard scalar product for
vectors $u,v \in {\mathbb R}^N$.

\begin{proposition}
Let $v_1, ... , v_N \in V$. The real $N \times N$ matrix $G:=\{g(v_h,v_j) \}$ is positive semidefinite and therefore ${\rm det}\{g(v_h,v_j) \} \geq 0$.
\end{proposition}
\begin{proof}
Let $x:=(x_1,...,x_N)\in {\mathbb R}^N$. We have
$$
0 \leq g\big(\sum_h x_hv_h,  \sum_h x_hv_h \big)=\sum_{h,j}
x_hx_jg(v_h,v_j)= \langle x, G(x) \rangle.
$$
\end{proof}

Motivated by the case $(V, g(\cdot,\cdot))=({\mathbb R}^N, \langle
\cdot, \cdot \rangle)$ one can give the following definition.

\begin{definition}
$$
{\rm Vol}^g(v_1, ... , v_N):=\sqrt{{\rm det}\{g(v_h,v_j) \}}.
$$
\end{definition}

\begin{remark}
\end{remark}

i) Obviously,
$$
{\rm Vol}^g(v_1, ... , v_N) \geq 0,
$$
where the equality holds if and only if $v_1, ... , v_N \in V$ are
linearly dependent.

ii) If the inner product depends on a further parameter so that
$g(\cdot,\cdot)=g_{\rho}(\cdot,\cdot)$, we write ${\rm
Vol}_{\rho}^g(v_1, ... , v_N)={\rm Vol}^g(v_1, ... , v_N)$.

iii) In the case of a probability space $(V,g_{\rho}(\cdot,\cdot))=({\cal L}^2_{\mathbb
R}(\Omega, {\cal G},\rho), {\rm Cov}_{\rho}(\cdot,\cdot))$ the
number ${\rm Vol}_{\rho}^{\rm Cov}(A_1, ... , A_N)^2$ is known as
the {\it generalized variance} of the random vector $(A_1, ... ,
A_N)$.

\bigskip

In what follows we move to the noncommutative case. Here
$A_1,...A_N$ are self-adjoint matrices, $\rho$ is a (faithful)
density matrix and $g(\cdot,\cdot)={\rm Cov}_\rho(\cdot,\cdot)$ has
been defined in  (\ref{cov}). By ${\rm Vol}_{\rho}^f$ we denote the
volume associated to the quantum Fisher information $\langle \cdot,\cdot\rangle_{\rho,f}$ given by the
(regular) normalized symmetric operator monotone function $f$.

\begin{definition}
\end{definition}
The function
$$
I_{\rho}^f(A):=\frac{f(0)}{2}{\rm Vol}_{\rho}^f(i[\rho,A])=\frac{f(0)}{2} \langle i[\rho,A],i[\rho,A] \rangle_{\rho,f}
$$
is known as the {\it metric adjusted skew information} or {\it $f$--information} (see \cite{Hansen:2006} \cite{GibiliscoImparatoIsola:2007}).

Let $N \in {\mathbb N}$, $ f \in {\cal F}_{op}^{\, r}$, $\rho \in {\cal D}^1_n$ and $A_1,...,A_N \in M_{n,sa}$ be arbitrary.
We shall prove in Section \ref{m} the following results.

\begin{theorem}

\label{main}
\begin{equation}
{\rm Vol}_{\rho}^{{\rm Cov}}(A_1, ... , A_N) \geq \left(
\frac{f(0)}{2}\right)^{\frac{N}{2}}{\rm Vol}_{\rho}^f(i[\rho,A_1],
... ,i[\rho,A_N]).  \label{con1}\end{equation}
\end{theorem}

\begin{theorem}

\label{equality}
The above inequality
is an equality if and only if ${A_1}_0, ... , {A_N}_0$ are linearly dependent.
\end{theorem}

\begin{theorem}

\label{monot} Fix $N \in {\mathbb N}$,  $\rho \in {\cal D}^1_n$ and $A_1,...,A_N \in M_{n,sa}$. Define for $f\in\mathcal F_{op}^{\, r}$
$$
V(f):=\left( \frac{f(0)}{2}\right)^{\frac{N}{2}}{\rm Vol}_{\rho}^f(i[\rho,A_1], ... ,i[\rho,A_N]).
$$
Then, for any $f,g\in\mathcal F_{op}^{\, r}$
$$
{\tilde f} \leq {\tilde g} \quad \Longrightarrow \quad V(f) \geq V(g).
$$
\end{theorem}

\begin{remark}

The inequality
$$
{\rm det} \{ {\rm Cov}_{\rho}(A_h,A_j) \} \geq
{\rm det} \left\{ {\rm Cov}_{\rho}(A_h,A_j)- {\rm Tr}(m_{\tilde f}(L_{\rho},R_{\rho})((A_h)_0)(A_j)_0)\right\}.
$$
makes sense also for not faithful states and it is true by continuity as a consequence of Theorem \ref{main}.

\end{remark}

Because of Proposition
\ref{puri} one has (by an obvious extension of the definition) the following result.

\begin{proposition}
If $\rho$ is a pure state then $ \forall N \in {\mathbb N}, \quad \forall f \in {\cal F}_{op}^{\, r}, \quad \forall A_1, ... , A_N \in M_{n,sa}$ one has
$$
{\rm Vol}_{\rho}^{{\rm Cov}}(A_1, ... , A_N)
=
\left( \frac{f(0)}{2}\right)^{\frac{N}{2}}{\rm Vol}_{\rho}^f(i[\rho,A_1], ... ,i[\rho,A_N]).
$$
\end{proposition}

\section{Proof of the main results} \label{m}

\begin{theorem} \label{main}

$$
{\rm Vol}_{\rho}^{{\rm Cov}}(A_1, ... , A_N)
\geq
\left( \frac{f(0)}{2}\right)^{\frac{N}{2}}{\rm Vol}_{\rho}^f(i[\rho, A_1], ... ,i[\rho, A_N])\qquad \qquad
\forall N \in {\mathbb N}^+, \quad \forall f \in {\cal F}_{op}^{\, r}.
$$

\end{theorem}
\begin{proof}
Theorem \ref{main} is equivalent to the following inequality
$$
{\rm det} \{ {\rm Cov}_{\rho}(A_h,A_j) \} \geq {\rm det} \left\{
\frac{f(0)}{2} \langle
i[\rho,A_h], i[\rho, A_j]\rangle_{\rho,f} \right\}.
$$
If $\rho$ and $A_1,...,A_N$ are fixed set
$$
F(f):={\rm det} \{ {\rm Cov}_{\rho}(A_h,A_j) \} - {\rm det} \left\{ \frac{f(0)}{2} \langle i[\rho, A_h], i[\rho, A_j]\rangle_{\rho,f} \right\}.
$$
Because of Theorem \ref{!} one has
$$
F(f)={\rm det} \{ {\rm Cov}_{\rho}(A_h,A_j) \} -
{\rm det} \left\{ {\rm Cov}_{\rho}(A_h,A_j)- {\rm Tr}(m_{\tilde f}(L_{\rho},R_{\rho})((A_h)_0)(A_j)_0)\right\}.
$$
Theorem \ref{main} is equivalent to
$$
F(f) \geq 0.
$$
From Proposition \ref{?}, we have

\begin{align*}
    {\rm Cov}_{\rho} (A_h,A_j)& = {\rm
    Re} \{{\rm Tr} (\rho (A_h)_0 (A_j)_0 \}= \frac{1}{2}
    \sum_{k,l}({\lambda}_k+{\lambda}_l) {\rm Re} \{\mathcal A_{kl}^hA_{lk}^j\} \\
    \frac{f(0)}{2}\langle i[\rho,A_h], i[\rho,A_j]\rangle_{\rho,f}
     &=\frac{1}{2} \sum_{k,l}({\lambda}_k+{\lambda}_l){\rm Re} \{\mathcal A_{kl}^h{\mathcal A}_{lk}^j\}  -
    \sum_{k,l} m_{\tilde f}(\lambda_k,\lambda_l){\rm Re} \{\mathcal A_{kl}^hA_{lk}^j\}  .
\end{align*}
We have
\[\begin{array}{rcl}
    F(f) &:=& \displaystyle\sum_{\sigma\in S_N}\sgn\sigma\left[ \prod_{j=1}^N
    {\rm Cov}_{\rho} (A_j,A_{\sigma(j)})-\prod_{j=1}^N\dfrac{f(0)}{2}\langle i[\rho,A_j],
    i[\rho,A_{\sigma(j)}]\rangle_{\rho,f}\right]\\[12pt]
    &:=&\displaystyle\sum_{\sigma\in S_N}\sgn\sigma\xi_\sigma,\end{array}\]
    where
\[ \begin{array}{rcl}
\xi_\sigma&=&\displaystyle\prod_{j=1}^N\sum_{k,l=1}^n\dfrac{\lambda_k+\lambda_l}2{\rm
Re} \{\mathcal A_{kl}^j\mathcal
A_{lk}^{\sigma(j)}\}-\prod_{j=1}^N\sum_{k,l=1}^n\left[\dfrac{\lambda_k+\lambda_l}2-m_{\tilde
f}(\lambda_k,\lambda_l)\right]{\rm Re} \{\mathcal
A_{kl}^j\mathcal A_{lk}^{\sigma(j)}\}.
\end{array}\]

From the Definition \ref{forse} we get (applying Proposition \ref{doppioscambio} to the case $X={\underline n}$)

\[ \begin{array}{rcl}
\xi_\sigma&=&\displaystyle\prod_{j=1}^N\sum_{k,l=1}^n\dfrac{\lambda_k+\lambda_l}2{\rm
Re} \{\mathcal A_{kl}^j\mathcal
A_{lk}^{\sigma(j)}\}-\prod_{j=1}^N\sum_{k,l=1}^n\left[\dfrac{\lambda_k+\lambda_l}2-m_{\tilde
f}(\lambda_k,\lambda_l)\right]{\rm Re} \{\mathcal
A_{kl}^j\mathcal A_{lk}^{\sigma(j)}\}
\\[12pt]
&=& \displaystyle\sum_{ \alpha, \beta\in\mathcal
C}\left\{\prod_{j=1}^N\dfrac{\lambda_{
\alpha_j}+\lambda_{\beta_j}}2{\rm Re} \{\mathcal A_{ \alpha_j
\beta_j}^j\mathcal A_{ \beta_j \alpha_j}^{\sigma(j)}\}
-\prod_{j=1}^N \left[\dfrac{\lambda_{ \alpha_j}+\lambda_{
\beta_j}}2-m_{\tilde f}(\lambda_{ \alpha_j},\lambda_{
\beta_j})\right]{\rm Re} \{\mathcal
A_{ \alpha_j \beta_j}^j\mathcal A_{ \beta_j \alpha_j}^{\sigma(i)}\}\right\}\\[12pt]
&=& \displaystyle\sum_{ \alpha, \beta\in\mathcal
C}\left\{\prod_{j=1}^N\dfrac{\lambda_{
\alpha_j}+\lambda_{\beta_j}}2\prod_{j=1}^N{\rm Re} \{\mathcal A_{
\alpha_j \beta_j}^j\mathcal A_{ \beta_j \alpha_j}^{\sigma(j)}\}
-\prod_{j=1}^N \left[\dfrac{\lambda_{ \alpha_j}+\lambda_{
\beta_j}}2-m_{\tilde f}(\lambda_{ \alpha_j},\lambda_{
\beta_j})\right]\prod_{j=1}^N{\rm Re} \{\mathcal
A_{ \alpha_j \beta_j}^j\mathcal A_{ \beta_j \alpha_j}^{\sigma(j)}\}\right\}\\[12pt]
&=& \displaystyle\sum_{ \alpha, \beta\in\mathcal
C}\left[\prod_{j=1}^N\dfrac{\lambda_{
\alpha_j}+\lambda_{\beta_j}}2-\prod_{j=1}^N\left( \dfrac{\lambda_{
\alpha_j}+\lambda_{ \beta_j}}2-m_{\tilde f}(\lambda_{
\alpha_j},\lambda_{ \beta_j})\right) \right]\prod_{j=1}^N{\rm Re} \{\mathcal
A_{ \alpha_j \beta_j}^j\mathcal A_{ \beta_j \alpha_j}^{\sigma(j)}\}\\[12pt]
&=& \displaystyle\sum_{ \alpha, \beta\in\mathcal C}H_{ \alpha,
\beta}\prod_{j=1}^N{\rm Re} \{\mathcal A_{ \alpha_j
\beta_j}^j\mathcal A_{ \beta_j \alpha_j}^{\sigma(j)}\}.
\end{array}\]
Hence, applying Proposition \ref{group} to the case $G=S^N$ , $X={\mathcal C} \times {\mathcal C}$ and $r(x):=r(\alpha,\beta):=H^f_{ \alpha, \beta}\det\mathcal B^{ \alpha, \beta}$ and Proposition \ref{sym} we get

\[\begin{array}{rcl}F(f)&=&\displaystyle\sum_{\sigma\in
S_N}\sgn\sigma\displaystyle\sum_{ \alpha, \beta\in\mathcal
C}H^f_{ \alpha, \beta}\prod_{i=1}^N{\rm Re} \{\mathcal
A_{ \alpha_i \beta_i}^i\mathcal A_{ \beta_i \alpha_i}^{\sigma(i)}\}\\[12pt]&=&
\displaystyle\sum_{ \alpha, \beta\in\mathcal
C}H^f_{ \alpha, \beta}\displaystyle\sum_{\sigma\in
S_N}\sgn\sigma\prod_{i=1}^N{\rm Re} \{\mathcal
A_{ \alpha_i \beta_i}^i\mathcal A_{ \beta_i \alpha_i}^{\sigma(i)}\}\\[12pt]
&=&\displaystyle \sum_{ \alpha, \beta\in\mathcal
C}H^f_{ \alpha, \beta}\det\mathcal B^{ \alpha, \beta}\\[12pt]
&=&\dfrac1{N!}\displaystyle \sum_{ \alpha, \beta\in\mathcal
C}H^f_{ \alpha, \beta}\sum_{\sigma\in S_N}\det\mathcal
B^{ \alpha_\sigma, \beta\sigma}\\[12pt]&=&\dfrac1{N!}\displaystyle \sum_{ \alpha, \beta\in\mathcal
C}H^f_{ \alpha, \beta}K_{ \alpha, \beta}.
\end{array}\]

By Corollary \ref{pos},
$H^f_{ \alpha, \beta}$ is strictly positive; on the other hand,
Lemma \ref{lem} ensures the nonnegativity of $K_{ \alpha, \beta}$,
so that we can conclude.
\end{proof}

\begin{theorem}
The inequality
$$
{\rm Vol}_{\rho}^{{\rm Cov}}(A_1, ... , A_N)
\geq
\left( \frac{f(0)}{2}\right)^{\frac{N}{2}}{\rm Vol}_{\rho}^f(i[\rho, A_1], ... ,i[\rho,A_N])\qquad \qquad
\forall N \in {\mathbb N}^+, \quad \forall f \in {\cal F}_{op}^{\, r}.
$$
is an equality if and only if ${(A_1)}_0, ... , {(A_N)}_0$ are
linearly dependent. \label{equ}\end{theorem}
\begin{proof}
Since
$$
{\rm Cov}_{\rho}(A_1,A_2)={\rm Tr}(\rho {(A_1)}_0 {(A_2)}_0) ={\rm
Cov}_{\rho}({(A_1)}_0, {(A_2)}_0)
$$
we have that
$$
{\rm Cov}_{\rho}(A_1,A_2)={\rm Cov}_{\rho}((A_1)_0 ,(A_2)_0).
$$

From this it follows

$$
{\rm Vol}_{\rho}^{{\rm Cov}}(A_1,\ldots , A_N) ={\rm
Vol}_{\rho}^{{\rm Cov}}({(A_1)}_0, \ldots , {(A_N)}_0)
$$
Therefore if ${(A_1)}_0, \ldots , {(A_N)}_0$ are linearly dependent
then

$$
0={\rm Vol}_{\rho}^{{\rm Cov}}({(A_1)}_0, \ldots , {(A_N)}_0)
={\rm Vol}_{\rho}^{{\rm Cov}}(A_1, ... , A_N) \geq \left(
\frac{f(0)}{2}\right)^{\frac{N}{2}}{\rm
Vol}_{\rho}^f(i[\rho,A_1], ... ,i[\rho,A_N]) \geq 0
$$
and we are done.

Conversely, suppose that $(A_1)_0,\ldots (A_N)_0$ are not linear
dependent; then we want to show that $F(f)>0$. Since for any
$\alpha,\beta\in\mathcal C$, $H_{\alpha,\beta}$ is strictly positive
and $K_{\alpha,\beta}$ is nonnegative, this is equivalent to prove that
$K_{\alpha,\beta}$ is not null for some $\alpha,\beta\in\mathcal C$.
Because of Lemma \ref{lem}, this is, in turn, equivalent to show
that $\det({\rm C}^{u(j)}\mathcal A_{\alpha_j,\beta_j}^i)$ is not
null for some $\alpha,\beta\in\mathcal C$ and $u\in \{0,1\}^{\underline N}$. This is a consequence of Corollary \ref{civuole}.

\end{proof}
\begin{theorem} \label{monot}
Define
$$
V(f):=\left( \frac{f(0)}{2}\right)^{\frac{N}{2}}{\rm
Vol}_{\rho}^f(i[\rho,A_1], ... ,i[\rho,A_N]).
$$
Then
$$
{\tilde f} \leq {\tilde g} \quad \Longrightarrow \quad V(f) \geq
V(g).
$$\label{monv}
\end{theorem}
\begin{proof} Because of Proposition \ref{posi} and Proposition \ref{mon}, one
has that
$$
\tilde{f} \leq \tilde{g} \quad \Longrightarrow \quad 0 <
H_{\alpha,\beta}^f \leq H_{\alpha,\beta}^f.
$$
Since $K_{\alpha,\beta} \geq 0$ does not depend on $f$ and
\[F(f)=\frac{1}{N!}\sum_{\alpha,\beta\in\mathcal C}H_{\alpha,\beta}^fK_{\alpha,\beta}\]
we get that  \[\quad 0
\leq F(f) \leq F(g).\]By definition of $F$, we obtain the thesis.
\end{proof}

\section{Relation with the standard uncertainty principle} \label{confronto}

\begin{theorem}(Hadamard inequality)

If $H \in M_{N,sa}$ is  positive semidifinite then
$$
\displaystyle {\rm det}(H) \leq \prod_{j=1}^N h_{jj}.
$$

\end{theorem}

\begin{theorem}
Let $f \in {\cal F}_{op}^{\, r}$. The inequality
$$
{\rm det} \left\{ \frac{f(0)}{2} \langle i[\rho, A_h],i[\rho,A_j] \rangle_{\rho,f}\right\}
\geq
{\rm det} \left\{ - \frac{i}{2} {\rm Tr}(\rho [A_h,A_j])\right\}
$$
is (in general) false for any $N=2m$.
\end{theorem}
\begin{proof}Let $n=N=2m$. By the Hadamard inequality it is enough to find
$A_1,..., A_N \in M_{N,sa}$  and a state $\rho \in {\cal D}^1_N$ such that
\begin{equation}
 \prod_{j=1}^N I_{\rho}^f(A_j) < {\rm det} \left\{ - \frac{i}{2} {\rm Tr}(\rho [A_h,A_j])\right\}.
\label{had}\end{equation} Let $\rho:={\rm
diag}(\lambda_1,...,\lambda_N)$, where
$\lambda_1<\lambda_2<\ldots<\lambda_N$. The aim is to construct
$A_1,\ldots A_N$  that are block-diagonal matrices, each matrix
consisting of exactly one non-null block equal to a $2\times 2$
Pauli matrix.

More precisely, given $h=2q+1$, where $q=0,\ldots N-1$, define the
Hermitian matrices $A_h$ and $A_{h+1}$ such that $(A_h)_{h\,
h+1}=i=(A_h^*)_{h+1\, h}$, $(A_{h+1})_{h\, h+1}=1=(A_{h+1})_{h+1\,
h}$ and $(A_h)_{kl}=(A_{h+1})_{kl}=0$ elsewhere.

Since the state $\rho$  is diagonal and $A_h$ are null diagonal matrices,
$A_h\equiv \mathcal A^h$, where $({\mathcal A}^h)_{kl}=\langle
(A_h)_0\phi_k,\phi_l\rangle$ is defined as in  Proposition \ref{?}. Therefore, say, if $h$ is odd one
obtains from Proposition \ref{fin}
\[\begin{array}{rcl}I_{\rho}^f(A_h) &=&\displaystyle
\frac{1}{2} \sum_{k,l}({\lambda}_k+{\lambda}_l)
    |{\cal A}^h_{kl}|^2-\sum_{k,l} m_{\tilde f}(\lambda_k,\lambda_l)
 |{\cal A}^h_{kl}|^2\\[12pt]&=&\lambda_h+\lambda_{h+1}-2m_{\tilde
 f}(\lambda_h,\lambda_{h+1})\\[12pt]&=&I_{\rho}^f(A_{h+1}).
\end{array}\] 

Suppose now that $h$ is odd and $h<k$. We have
\[\begin{array}{rcl} {\rm
Tr}(\rho [A_h,A_k])
&=&\displaystyle
\sum_{j,l,m}\rho_{jl}\big((A_h)_{lm}(A_k)_{mj}-(A_k)_{lm}(A_h)_{mj}\big)\\[12pt]
&=& \displaystyle\sum_{j,m}\lambda_j\big((A_h)_{jm}(A_k)_{mj}-(A_k)_{jm}(A_h)_{mj}\big)\\[12pt]
&=&\displaystyle\sum_{j,m}\lambda_j((A_h)_{jm}(A_k)_{mj}-(A_k)_{jm}(A_h)_{mj})\\[12pt]
&=&2i(\lambda_{h}-\lambda_{h+1})\delta_k^{h+1},
\end{array}\]
where $\delta_h^{k+1}$ denotes the Kronecker delta function. We have that ${\rm
Tr}(\rho [A_h,A_k])=-{\rm Tr}(\rho [A_k,A_h])$ and therefore,
$$
\left\{ - \frac{i}{2} {\rm Tr}(\rho [A_h,A_j])\right\}
=
\left(
\begin{array}{ccccc}
     0&  \lambda_1-\lambda_2& 0 & \ldots& 0\\
 \lambda_2-\lambda_1&  0& \lambda_2-\lambda_3&\ldots &0\\
0 & \lambda_3-\lambda_2& 0&\ldots & 0 \\
\ldots & \ldots &\ldots& \ldots & \ldots \\ 
0& \ldots& \ldots& 0 & \lambda_{N-1}-\lambda_N\\
0& 0& \ldots&  \lambda_{N}-\lambda_{N-1} & 0\\
\end{array}
\right)
$$
so that
\[{\rm det} \left\{ - \frac{i}{2} {\rm Tr}(\rho [A_h,A_j])\right\}=\prod_{ h <N, \, h=2q+1}(\lambda_{h+1}-\lambda_h)^2\]
Finally, since for any $f \in {\cal F}_{op}^{\, r}$ the function $m_{\tilde
 f}(\cdot,\cdot)$ is a mean, one has
$\lambda_h< m_{\tilde
 f}(\lambda_h,\lambda_{h+1})<\lambda_{h+1}$. This implies,  for any odd $h$,
 \[I_{\rho}^f(A_h) =I_{\rho}^f(A_{h+1})= \lambda_h+\lambda_{h+1}-2m_{\tilde
 f}(\lambda_h,\lambda_{h+1})<\lambda_{h+1}-\lambda_h,\]
 so that one can get (\ref{had}) by taking the product over all $h$.

\end{proof}

\begin{theorem}
Let $f \in {\cal F}_{op}^{\, r}$. The inequality
$$
{\rm det} \left\{ \frac{f(0)}{2} \langle i[\rho, A_h],i[\rho,A_j] \rangle_{\rho,f}\right\}
\leq
{\rm det} \left\{ - \frac{i}{2} {\rm Tr}(\rho [A_h,A_j])\right\}
$$
is (in general) false for any $N=2m$.
\end{theorem}
\begin{proof}
It suffices to find selfadjoint matrices $A_1,\ldots A_N$ which are pairwise commuting  but  not commuting with a given state $\rho$ and such that
$[\rho,A_1],\ldots [\rho, A_N]$ are linearly independent.

Consider a state of the form $\rho={\rm diag}(\lambda_1,...,\lambda_n)$ where the eigenvalues $\lambda_i$ are all distinct.

Let $A_1,\ldots A_N\in
\mathcal M_{n,sa}(\mathbb R)$ be $N$ linear independent symmetric
real matrices such that $(A_j)_{kk}=0$ for any $j=1,\ldots N$ and
$k=1,\ldots, n$. Note that the linear independence of $A_1,\ldots A_N$ implies the condition
 $n(n-1)/2\geq N$.

Obviously, $[A_j,A_m]=0$ for any $j,m=1,\ldots N$, while  a direct
computation shows that
\[\begin{array}{rcl}([\rho,A_j])_{kl}&=&\displaystyle\sum_{h=1}^n\rho_{kh}(A_j)_{hl}-\sum_{h=1}^n(A_j)_{kh}\rho_{hl}
\\[12pt]&=&(A_j)_{kl}(\lambda_{k}-\lambda_{l})\end{array}\]
Observe that also $[\rho,A_1]\ldots [\rho,A_N]$ are linear
independent. Suppose, in fact, that there exists a vector
$\alpha\in\mathbb R^N$ such that
\[\sum_{j=1}^N\alpha_j[\rho,A_j]\equiv 0,\]that is, for any $k,l=1,\ldots n$
\[0=\sum_{j=1}^N\alpha_j([\rho,A_j])_{kl}= (\lambda_{k}-\lambda_{l})\sum_{j=1}^N\alpha_j(A_j)_{kl}.\]
This implies that $\sum_{j}\alpha_j(A_j)_{kl}=0$, and hence
$\alpha\equiv 0$, because of the linear independence of $A_1,\ldots
A_N$.

\end{proof}

\section{Appendix A: combinatorics} \label{comb}

Set $\underline n := \{1,...,n \}$. Moreover define
\[\mathcal C:=\underline{n}^{\underline N}=\{(x_1,\ldots, x_N):
x_i\in\{1,\ldots n\},\ i=1,\ldots N\}.\]

One can prove the following result.

\begin{proposition} \label{scambio}
For a finite set $X \subset \mathbb N$ and $N \in {\mathbb N}^+$ one has
$$
\displaystyle\prod_{j=1}^N \sum_{k\in X} Q^k_j
=
\sum_{u\in X^{\underline N}}\displaystyle\prod_{j=1}^N Q^{u(j)}_j
$$
\end{proposition}
For $z \in \mathbb C$
we shall introduce the operator
\[{\rm C}^k(z):=\left\{\begin{array}{rcl}{\rm Re}(z)&\text{if}&k=0,\\[12pt]{\rm Im}(z)&\text{if}&k=1.\end{array}\right.\]

Taking $X=\{0,1\}$ and $Q^k_j=C^k(z_j)C^k(w_j)$ in Proposition \ref{scambio} we get

\begin{corollary} \label{idea}If $z_j,w_j \in \mathbb C$ then
$$
\displaystyle\prod_{j=1}^N \left( \sum_{k\in \{0,1\}} {\rm C}^k(z_j){\rm C}^k(w_j) \right)
=
\sum_{u\in \{0,1\}^{\underline N}}\left( \displaystyle\prod_{j=1}^N {\rm C}^{u(j)}(z_j){\rm C}^{u(j)}(w_j) \right).
$$
\end{corollary}

With similar arguments it is possible to prove the following result.

\begin{proposition} \label{doppioscambio}
For a finite set $X \subset \mathbb N$ and $N \in {\mathbb N}^+$ one has
$$
\displaystyle\prod_{j=1}^N \left( \sum_{k,l\in X} Q^j_{kl} \right)
=
\sum_{\alpha,\beta\in X^{\underline N}}\left( \displaystyle\prod_{j=1}^N Q^{j}_{\alpha(j) \beta(j)} \right)
$$
\end{proposition}

The following result is obvious.
\begin{proposition}
Let $X$ be a finite set and $g:X \to X$ a bijection. For any function $r:X \to \mathbb R$ one has

\[\sum_{x\in X}r(x)=
\sum_{x\in X}r(g(x)).\]

\end{proposition}

From the above result one obtains the following

\begin{proposition}\label{group}
Let $X$ be a finite set and let $G$ be a group of bijections $g:X \to X$. For any function $r:X \to \mathbb R$ one has

\[\sum_{x\in X}r(x)=\frac{1}{\sharp(G)}\sum_{x\in X} \sum_{g\in G} r(g(x)).\]

\end{proposition}

We denote by $S^N$ the symmetric group of order $N$.

\begin{example}
The set $X:=\{0,1\}^{\underline N}$ can be identified with the power set of $\underline N$. If $u\in\{0,1\}^{\underline N}$ and $\sigma\in S_N$ the $\sigma$ can be seen as a bijection
$\sigma :X \to X$ defining $\sigma(u):=u \circ \sigma$.
\end{example}

From the above considerations we get the following Lemma.

\begin{lemma} \label{sum} For any function $r:\{0,1\}^{\underline N}\to \mathbb R$ and for any $\sigma \in S_N$ one has
\[\sum_{u\in \{0,1\}^{\underline N}}r(u)=
\sum_{u\in \{0,1\}^{\underline N}}r(\sigma(u)).\]
\end{lemma}

\begin{remark}\label{righe}
If $E=\{E_{jk}\}$ and $E(\sigma):=\{E_{\sigma(j)k}\}$ one has
$$
{\rm det}(E(\sigma))={\rm sgn}(\sigma) {\rm det}(E).
$$
\end{remark}

\section{Appendix B: the function $H$} \label{H}

Let $\mathbb R_+ := (0,+\infty)$ and $\mathbf x=(x_1,...,x_N), \mathbf y=(y_1,...,y_N) \in \mathbb R_+^N$.

In the sequel we need to study the following function.

\begin{definition}
For any $f \in {\cal F}_{op}^{\, r}$,
set
\[
H^{ f}(\mathbf{x},\mathbf
{y}):=\prod_{j=1}^N\dfrac{x_j+y_j}2-\prod_{j=1}^N\left(\dfrac{x_j+y_j}2-m_{\tilde
f}(x_j,{y_j})\right)
\]
\end{definition}
\begin{proposition}
For any $\mathcal F\in\mathcal F_{op}^{\, r}$, $\mathbf x, \mathbf
y\in\mathbb R_+^N$, $$H^f(\mathbf x,\mathbf y)>0.$$
\label{posi}\end{proposition}
\begin{proof} Since for any $\mathbf x, \mathbf y\in\mathbb
R_+^N$,
\[0<m_{\tilde f}(x_j,y_j)\le \frac{x_j+y_j}2, \quad j=1,\ldots N,\]
we have
\[\prod_{j=1}^N\left(\dfrac{x_j+y_j}2-m_{\tilde
f}(x_j,{y_j})\right) <\prod_{j=1}^N\dfrac{x_j+y_j}2,\] so that we
can conclude.
\end{proof}
\begin{proposition}
$$
\tilde{f} \leq \tilde{g}
$$
$$
\Downarrow
$$
$$
H^{f}(\mathbf x,\mathbf y) \leq H^g(\mathbf x,\mathbf y) \qquad
\qquad \forall \mathbf x,\mathbf y\in\mathbb R_+^N.
$$\label{mon}
\end{proposition}

\begin{proof}

Since for any $x,y>0$
\begin{equation}
\frac{x+y}2-m_{\tilde f}(x,y)= \frac{(x-y)^2}{2y} \cdot
\frac{f(0)}{f(\frac{x}{y}) }, \label{f}\end{equation} we have
\[
H^{ f}(\mathbf{x},\mathbf
{y}):=\prod_{j=1}^N\dfrac{x_j+y_j}2-\prod_{j=1}^N\left(\frac{(x_j-y_j)^2}{2y_j}
\cdot \frac{f(0)}{f(\frac{x_j}{y_j}) },\right) .\]

Because of Proposition \ref{min} we have

$$
\tilde{f} \leq \tilde{g} \Rightarrow \frac{f(0)}{f(t)} \geq
\frac{g(0)}{g(t)} > 0 \qquad \qquad \forall t >0;
$$
hence, we obtain
$$
H^{ f}(\mathbf x,\mathbf y) \leq H^g(\mathbf x,\mathbf y) \qquad
\qquad \forall \mathbf x,\mathbf y\in\mathbb R_+^N
$$
by elementary computations.
\end{proof}

\begin{corollary} \label{mon2}
For any $f \in {\cal F}_{op}$,

\[0<H^{SLD}(\mathbf x,\mathbf y)\le H^{ f}(\mathbf x,\mathbf y)\leq \frac{1}{2^N}\prod_{j=1}^N(x_j+y_j)\qquad \qquad
\forall \mathbf x,\mathbf y\in\mathbb R_+^N .\]\label{pos}
\end{corollary}

\begin{definition} \label{forse} Fix $(\lambda_1,...,\lambda_n) \in {\mathbb R}^n_+$.
Given $ \alpha,
\beta\in\mathcal C={\underline n}^{\underline N}$, let $H_{ \alpha, \beta}^f:=H^f(\lambda_{
\alpha},\lambda_{ \beta})$, where $\lambda_{ \alpha}:=(\lambda_{
\alpha_1},\ldots ,\lambda_{ \alpha_N})$, $\lambda_{
\beta}:=(\lambda_{ \beta_1},\ldots ,\lambda_{ \beta_N})$.
\end{definition}

\begin{proposition} \label{sym}
For all $\sigma \in S^N$ one has
$$
H_{ \alpha(\sigma), \beta(\sigma)}^f=H_{ \alpha, \beta}^f.
$$
\end{proposition}
\begin{proof}
Left to the reader
\end{proof}
\section{Appendix C: the function $K$} \label{K}

In order to prove the main result of this paper, we introduce some
notations. Let $\left\{\varphi_i\right\}$ be a complete orthonormal
base composed of eigenvectors of $\rho$, and $\{ {\lambda}_i \}$ the
corresponding eigenvalues. As in Proposition \ref{?} set
\[\mathcal A_{kl}^j:=\langle(A_j)_0\ {\varphi}_{ k}|{\varphi}_l\rangle \qquad \qquad  j=1,...,N; \quad  \quad k,l=1,...,n.\]

Note that since the $A_j$ are selfadjoint one has ${\mathcal A}_{kl}^j= \overline {{\mathcal A}_{lk}^j}$ namely
$$
{\rm Re}({\mathcal A}_{kl}^j)={\rm Re}({\mathcal A}_{lk}^j) \qquad \qquad {\rm Im}({\mathcal A}_{kl}^j)=-{\rm Im}({\mathcal A}_{lk}^j)
$$
Since
$$
{\rm Re}(zw)={\rm Re}(z){\rm Re}(w)-{\rm Im}(z){\rm Im}(w)
$$
we have
\begin{lemma} \label{deve}
$$
{\rm Re}({\mathcal A}_{kl}^j{\mathcal A}_{lk}^m)=
{\rm Re}({\mathcal A}_{kl}^j){\rm Re}({\mathcal A}_{kl}^m)+{\rm Im}({\mathcal A}_{kl}^j){\rm Im}({\mathcal A}_{kl}^m)
$$
\end{lemma}

If $\alpha, \beta \in {\mathcal C}={\underline n}^{\underline N}$ and $\sigma \in S^N$ we define a $N \times N$ matrix $\mathcal B^{ \alpha_\sigma, \beta_\sigma}$ setting

\[\left(\mathcal B^{ \alpha_\sigma, \beta_\sigma}\right)_{hj}:={\rm
Re}\{ \mathcal A_{ \alpha_{\sigma(h)}, \beta_{\sigma(h)}}^h \mathcal
A_{\beta_{\sigma(h)},\alpha_{\sigma(h)}}^{j}\}. \qquad \qquad  h,j=1,...,N; \quad  \quad \alpha_{\sigma(h)},\beta_{\sigma(h)}=1,...,n\]

When $\sigma:=I$
is the identity in $S_N$, we shall simply denote by $\mathcal A_{
\alpha, \beta}$ and $\mathcal B^{ \alpha, \beta}$ the corresponding
matrices.

\begin{definition}
$$
K_{\alpha,\beta}:= K_{\alpha,\beta}(\rho;A_1,...,A_N):=\sum_{\sigma\in S_N}\det \left( \mathcal B^{ \alpha_\sigma, \beta_\sigma} \right)
$$
\end{definition}

\begin{definition}
If $u \in \{0,1 \}^{\underline N}$ and $\alpha,\beta \in {\underline n}^{\underline N}$ we define an $N \times N$ matrix setting
$$
D(u;\alpha,\beta):=\{D(u;\alpha,\beta)_{hj} \}:=\{{\rm C}^{u(j)} {\mathcal A}^h_{\alpha_j \beta_j}\} \qquad h,j=1,...,N
$$
\end{definition}

\begin{proposition} We have
\[
K_{\alpha,\beta}=
\displaystyle\sum_{u\in\{0,1\}^N} \det(D(u;\alpha,\beta) )^2 \geq 0.\]\label{lem}
so that $K_{\alpha,\beta} \geq 0$.
\end{proposition}
\begin{proof}

Applying:

i) Lemma \ref{deve};

ii) Corollary \ref{idea};

iii) Lemma \ref{sum}.

to the function

$$r(u)=r_{\sigma,\tau}(u):=\displaystyle\prod_{j=1}^N{\rm
C}^{u(j)}\mathcal A_{ \alpha_{\sigma(j)},
\beta_{\sigma(j)}}^j\mathcal \displaystyle{\rm C}^{u(j)}\mathcal A_{
\alpha_{\sigma(j)}, \beta_{\sigma(j)}}^{\tau(j)},$$

we get

\[\begin{array}{rcl}K_{\alpha,\beta}&=&\displaystyle\sum_{\sigma\in S_N}\det\mathcal
B^{ \alpha_\sigma, \beta_\sigma}\\[12pt]
&=&\displaystyle\sum_{\sigma\in
S_N}\displaystyle\sum_{\tau\in S_N}\sgn\tau\prod_{j=1}^N\mathcal
({\cal B}^{ \alpha_\sigma, \beta_\sigma})_{j,\tau(j)}
\\[12pt]
&=&\displaystyle\sum_{\sigma\in
S_N}\displaystyle\sum_{\tau\in S_N}\sgn\tau\prod_{j=1}^N{\rm Re}
\{\mathcal A_{ \alpha_{\sigma(j)}, \beta_{\sigma(j)}}^j\mathcal
A_{\beta_{\sigma(j)},\alpha_{\sigma(j)}}^{\tau(j)}\}
\\[12pt]
&=&
\displaystyle\sum_{\sigma\in S_N}\displaystyle\sum_{\tau\in
S_N}\sgn\tau\prod_{j=1}^N\left({\rm Re}\mathcal A_{
\alpha_{\sigma(j)}, \beta_{\sigma(j)}}^j{\rm Re}\mathcal A_{
\alpha_{\sigma(j)}, \beta_{\sigma(j)}}^{\tau(j)}+{\rm Im}{\mathcal A}_{
\alpha_{\sigma(j)}, \beta_{\sigma(j)}}^j{\rm Im}\mathcal A_{
\alpha_{\sigma(j)}, \beta_{\sigma(j)}}^{\tau(j)}\right)
\\[12pt]
&=&
\displaystyle\sum_{\sigma\in S_N}\displaystyle\sum_{\tau\in
S_N}\sgn\tau\prod_{j=1}^N\left( \sum_{u\in \{0,1\}}{\rm C}^u\mathcal A_{
\alpha_{\sigma(j)}, \beta_{\sigma(j)}}^j
\cdot
{\rm C}^u\mathcal A_{
\alpha_{\sigma(j)}, \beta_{\sigma(j)}}^{\tau(j)}\right)
\\[12pt]
&=&\displaystyle\sum_{\sigma\in
S_N}\displaystyle\sum_{\tau\in
S_N}\sgn\tau\sum_{u\in\{0,1\}^{\underline N}}\displaystyle\prod_{j=1}^N{\rm
C}^{u(j)}\mathcal A_{ \alpha_{\sigma(j)},
\beta_{\sigma(j)}}^j\mathcal \displaystyle{\rm C}^{u(j)}\mathcal A_{
\alpha_{\sigma(j)}, \beta_{\sigma(j)}}^{\tau(j)}
\\[12pt]
&=&\displaystyle\sum_{\sigma\in
S_N}\displaystyle\sum_{\tau\in
S_N}\sgn\tau\sum_{u\in\{0,1\}^{\underline N}}\displaystyle\prod_{j=1}^N{\rm
C}^{u(\sigma(j))}\mathcal A_{ \alpha_{\sigma(j)},
\beta_{\sigma(j)}}^j\mathcal \displaystyle{\rm
C}^{u(\sigma(j))}\mathcal A_{ \alpha_{\sigma(j)},
\beta_{\sigma(j)}}^{\tau(j)},\end{array}\]

Hence, by Remark \ref{righe}
\[\begin{array}{rcl}
K_{\alpha,\beta}
&=&\displaystyle\sum_{\sigma\in S_N}\det\mathcal
B^{ \alpha_\sigma, \beta_\sigma}\\[12pt]
&=&\displaystyle\sum_{u\in\{0,1\}^N}\sum_{\sigma\in
S_N}\displaystyle\sum_{\tau\in
S_N}\sgn\tau\displaystyle\prod_{j=1}^N{\rm C}^{u(\sigma(j))}\mathcal
A_{ \alpha_{\sigma(j)}, \beta_{\sigma(j)}}^j
\displaystyle\prod_{h=1}^N{\rm C}^{u(\sigma(h))}\mathcal A_{
\alpha_{\sigma(h)}, \beta_{\sigma(h)}}^{\tau(h)}
\\[12pt]
&=&\displaystyle\sum_{u\in\{0,1\}^N} \displaystyle\sum_{\sigma\in
S_N}\sgn\sigma\prod_{j=1}^N{\rm C}^{u(\sigma(j))}\mathcal A_{
\alpha_{\sigma(j)}, \beta_{\sigma(j)}}^j \sgn\sigma\displaystyle\sum_{\tau\in
S_N}\sgn\tau\prod_{h=1}^N{\rm C}^{u(\sigma(h))}\mathcal A_{
\alpha_{\sigma(h)}, \beta_{\sigma(h)}}^{\tau(h)}
\\[12pt]
&=&\displaystyle\sum_{u\in\{0,1\}^N} \left( \displaystyle\sum_{\sigma\in
S_N}\sgn\sigma\prod_{j=1}^N{\rm C}^{u(\sigma(j))}\mathcal A_{
\alpha_{\sigma(j)}, \beta_{\sigma(j)}}^j \right) \left( \sgn\sigma \det\left\{{\rm
C}^{u(\sigma(h))}\mathcal A_{\alpha_{\sigma(h)},
\beta_{\sigma(h)}}^{j}\right\} \right)
\\[12pt]
&=&\displaystyle\sum_{u\in\{0,1\}^N}
\det\left\{{\rm C}^{u(j)}\mathcal A_{\alpha_{j},
\beta_{j}}^{h}\right\} \det\left\{{\rm C}^{u(h)}\mathcal
A_{\alpha_{h}, \beta_{h}}^{j}\right\}
\\[12pt]
&=&\displaystyle\sum_{u\in\{0,1\}^N}
\left(\det\left\{{\rm C}^{u(j)}\mathcal A_{\alpha_{j},
\beta_{j}}^{h}\right\}\right)^2
\\[12pt]
&=&\displaystyle\sum_{u\in\{0,1\}^N} \det(D(u;\alpha,\beta) )^2.
\end{array}\]
\end{proof}

\begin{lemma} \label{bravo}
If $\mathcal A^1, \ldots \mathcal A^N \in M_{n,sa}$ are linearly independent then there exist $\alpha,\beta\in\mathcal C$ and $u\in \{0,1\}^{\underline N}$ such that
$$
\det\{{\rm
C}^{u(j)}\mathcal A_{\alpha_j,\beta_j}^h \}\neq 0.
$$

\end{lemma}
\begin{proof}
Note that the independence hypothesis implies $N \leq {\rm dim}_{\mathbb R}(M_{n,sa})=n^2$. Therefore the $N \times n^2$ matrix

\[
\left(
\begin{array}{cccccccccc}
     \mathcal A^1_{11}& \ldots & \mathcal A^1_{1n} & \mathcal A^1_{21}& \ldots &\mathcal A^1_{2n}& \ldots&\mathcal A^1_{n1}& \ldots & \mathcal A^1_{nn} \\
  \mathcal A^2_{11}& \ldots & \mathcal A^2_{1n} & \mathcal A^2_{21}& \ldots &\mathcal A^2_{2n}& \ldots&\mathcal A^2_{n1}& \ldots & \mathcal A^2_{nn} \\
\ldots& \ldots & \ldots & \ldots& \ldots & \ldots& \ldots&\ldots& \ldots & \ldots \\
  \mathcal A^N_{11}& \ldots & \mathcal A^N_{1n} & \mathcal A^N_{21}& \ldots &\mathcal A^N_{2n}& \ldots&\mathcal A^N_{n1}& \ldots & \mathcal A^N_{nn} \\
  \end{array}
\right)
\]
has rank $N$ because it has $N$ independent rows. This means that there exists $N$ columns that are linearly independent and this is equivalent to say that there exists $\alpha,\beta\in\mathcal C$ such that the matrix

\[
\left(
\begin{array}{cccc}
     \mathcal A^1_{\alpha_1\beta_1}&  \mathcal A^1_{\alpha_2\beta_2}& \ldots &\mathcal A^1_{\alpha_N\beta_N}\\
 \mathcal A^2_{\alpha_1\beta_1}&  \mathcal A^2_{\alpha_2\beta_2}& \ldots &\mathcal A^2_{\alpha_N\beta_N}\\
\ldots & \ldots & \ldots & \ldots \\
 \mathcal A^N_{\alpha_1\beta_1}&  \mathcal A^N_{\alpha_2\beta_2}& \ldots &\mathcal A^N_{\alpha_N\beta_N}\\
\end{array}
\right)
\]
has rank $N$. This implies that the $N \times 2N$ matrix

\[
\left(
\begin{array}{ccccccc}
     {\rm Re}\mathcal A^1_{\alpha_1\beta_1}& {\rm Im}\mathcal A^1_{\alpha_1\beta_1} &
{\rm Re}\mathcal A^1_{\alpha_2\beta_2}& {\rm Im}\mathcal A^1_{\alpha_2\beta_2}
& \ldots & {\rm Re}\mathcal A^1_{\alpha_N\beta_N}& {\rm Im}\mathcal A^1_{\alpha_N\beta_N}\\
{\rm Re}\mathcal A^2_{\alpha_1\beta_1}& {\rm Im}\mathcal A^2_{\alpha_1\beta_1} &
{\rm Re}\mathcal A^2_{\alpha_2\beta_2}& {\rm Im}\mathcal A^2_{\alpha_2\beta_2}
& \ldots & {\rm Re}\mathcal A^2_{\alpha_N\beta_N}& {\rm Im}\mathcal A^2_{\alpha_N\beta_N}\\
\ldots& \ldots & \ldots & \ldots& \ldots & \ldots& \ldots\\
{\rm Re}\mathcal A^N_{\alpha_1\beta_1}& {\rm Im}\mathcal A^N_{\alpha_1\beta_1} &
{\rm Re}\mathcal A^N_{\alpha_2\beta_2}& {\rm Im}\mathcal A^N_{\alpha_2\beta_2}
& \ldots & {\rm Re}\mathcal A^N_{\alpha_N\beta_N}& {\rm Im}\mathcal A^N_{\alpha_N\beta_N}\\
 \end{array}
\right)
\]
has  rank $N$  because it has $N$ indipendent rows. Therefore this matrix must have also $N$ independent columns. This last assertion it is equivalent to the desired conclusion

\end{proof}
\begin{corollary}\label{civuole}

If $(A_1)_0,...,(A_N)_0\in M_{n,sa}$ are linear independent then there exist   $\alpha, \beta \in \mathcal C$ and $u \in \{0,1 \}^{\underline N}$ such that
$$
\det(D(u;\alpha,\beta))=\det\{{\rm C}^{u(j)}\mathcal A_{\alpha_j,\beta_j}^h\} \not=0.
$$

\end{corollary}

\begin{proof}
By
definition of $\mathcal A^j$, $j=1,\ldots, N$, observe that the hypothesis of linear independence of
$$
(A_1)_0,\ldots , (A_N)_0
$$
implies the linear independence
of
$$
\mathcal A^1, \ldots \mathcal A^N.
$$
 Hence, by Lemma \ref{bravo} there exist
$\alpha,\beta\in\mathcal C$  and $u\in \{0,1\}^{\underline N}$ such that
$$
\det(D(u;\alpha,\beta))
=
\det\{{\rm
C}^{u(j)}\mathcal A_{\alpha_j,\beta_j}^h \}\neq 0.
$$

\end{proof}

\end{document}